\documentclass[aps]{revtex4}
\usepackage{amssymb}
\usepackage{color}
\usepackage[colorlinks=true,linkcolor=red]{hyperref}
\usepackage{hyperref}
\begin{document}
\title{Gravitational Localization of Matters in 6D}
\author{Pavle  Midodashvili}
\email{midodashvili@hotmail.com}
\affiliation{Department of
Physics, Tskhinvali State University, 2 Besiki Str., Gori 383500,
GEORGIA}
\author{Levan  Midodashvili}
\email{Levmid@hotmail.com} \affiliation{Department of Physics,
Tskhinvali State University, 2 Besiki Str., Gori 383500, GEORGIA}
\date{August 5, 2003}
\begin{abstract}
We present a new 3-brane solution to Einstein's equations in
(1+5)-spacetime with a negative bulk cosmological constant. This
solution is a stringlike defect solution with decreasing scale
function approaching a finite non-zero value in the radial
infinity. It is shown that all local fields are localized on the
brane only through the gravitational interaction.
\end{abstract}

\maketitle

It is believed that the idea of extra dimensions would be one of
the most attractive ideas concerning unification of gauge fields
with general relativity. Large extra dimensions offer an
opportunity for a new solutions to old problems (smallness of
cosmological constant, the origin of the hierarchy problem, the
nature of flavor, etc.). This idea has been much investigated
since the appearance of papers \cite{RSh-Visser-RS-ADD-Gogber}. In
such theories our world can be associated with a $3$-brane,
embedded in a higher-dimensional space-time with non-compact extra
dimensions and non-factorizable geometry. In this scenario it is
assumed that all the matter fields are constrained to live on the
$3$-brane and the corrections to four dimensional Newton's gravity
low from bulk gravitons are small for macroscopic scales. But this
models still need some natural mechanism of localization of known
particles on the brane. The questions of matter localization on
the brane has been investigated in various papers
\cite{mg1,mg2,Chacko-Oda-GRSh-GMSh-Giovan-R_DSh-GSH-CP-CK-BG}. In
our opinion the localizing force must be universal for all types
of 4- dimensional matter fields. In our world the gravity is known
to be the unique interaction which has universal coupling with all
matter fields. So if extended extra dimensions exist, it is
natural to assume that trapping of matter on the brane has a
gravitational nature. It is of interest that recently in
\cite{GogberSigleton} it was found the model in the brane world
where all the local bulk fields (ranging from the spin 0 scalar
field to the spin 2 gravitational field) are localized on the
3-brane only by the universal interaction, i.e., the gravity. The
solution is found in (1+5)-spacetime for the positive bulk
cosmological constant $\Lambda > 0$ and has increasing scale
factor $\phi (r)$ asymptotically approaching finite value at the
radial infinity. Recently in \cite{Oda1}  this solution was
extended to the case of a general (p-1)-brane model with
codimension $n$ in general $D=p+n$ space-time dimensions.

In this article we introduce the new 3-brane solution in
(1+5)-spacetime for the negative cosmological constant $\Lambda <
0$ which has decreasing scale factor $\phi (r)$ asymptotically
approaching finite non-zero value at the radial infinity. We
explicitly show that whole local fields (spin $0$ scalar field,
spin ${\raise0.7ex\hbox{$1$} \!\mathord{\left/
 {\vphantom {1 2}}\right.\kern-\nulldelimiterspace}
\!\lower0.7ex\hbox{$2$}}$ spinor field, spin 1 gauge field, spin
${\raise0.7ex\hbox{$1$} \!\mathord{\left/
 {\vphantom {3 2}}\right.\kern-\nulldelimiterspace}
\!\lower0.7ex\hbox{$2$}}$ gravitino field and spin $2$
gravitational field) as well as totally antisymmetric tensor
fields are localized on the $3$-brane by the gravity.

Let us begin with the details of our solution. In 6D the Einstein
equations with a bulk cosmological constant $\Lambda$ and
stress-energy tensor $T_{AB}$  \begin{equation}
\label{6Dequations}R_{AB}  - \frac{1}{2}g_{AB} R = \frac{1}{{M^4
}}(\Lambda g_{AB}  + T_{AB} )\end{equation} can be derived from
the whole action of the gravitating system
\begin{equation}S = \int {d^6 x\sqrt { - g} \left[ {\frac{{M^4 }}{2}\left( {R +
2\Lambda } \right) + L} \right]}\end{equation}  $R_{AB}$ , $R$,
$M$ and $L$ are respectively the Ricci tensor, the scalar
curvature, the fundamental scale and the Lagrangian of  matter
fields (including brane). All of these physical quantities refer
to $(1+5)$- space with signature $(+ - ... -)$, capital Latin
indices run over $A,B,...=0,1,2,3,5,6$. Suppose that the equations
(\ref{6Dequations})  admit a solution that is consistent with
four-dimensional Poincar\'{e} invariance. Introducing for the
extra dimensions the polar coordinates $(r,\theta )$, where $0 \le
r <  + \infty$ , $0 \le \theta  < 2\pi$ , the six-dimensional
metric satisfying this ansatz we can choose in the form
\cite{mg2,GogberSigleton} :
\begin{equation}\label{ansatzA}ds^2  = \phi ^2 \left( r \right)\eta _{\alpha \beta } \left(
{x^\nu  } \right)dx^\alpha  dx^\beta   - g\left( r \right)\left(
{dr^2  + r^2 d\theta ^2 } \right) ,\end{equation} where small
Greek indices $\alpha ,\beta ,... = 0,1,2,3$ numerate coordinates
and physical quantities in four-dimensional space, the functions
$\phi (r )$ and $g(r)$ depend only on  $r$  and are cylindrically
symmetric in the extra-space, the metric signature is $(+ - ...
-)$.The function $g(r)$ must be positive to fix the signature of
the metric (\ref{ansatzA}).

The source of the brane  is described by a stress-energy tensor
$T_{AB}$ also cylindrically symmetric in the extra-space. Its
nonzero components we choose in the form
\begin{equation}\label{branetensor}T_{\alpha \beta }  =  - g_{\alpha \beta } F_0 \left( r \right),\ \
T_{ij}  =  - g_{ij} F\left( r \right) , \end{equation} where we
have introduced two source functions $F_0$  and $F$, which depend
only on the radial coordinate $r$.

By using cylindrically symmetric metric ansatz (\ref{ansatzA})
 and stress-energy tensor (\ref{branetensor}),  the Einstein equations become
\begin{equation}\label{4dpartA}3\frac{{\phi ''}}{\phi } + 3\frac{{\phi '^2 }}{{\phi ^2 }} +
3\frac{{\phi '}}{{r\phi }} + \frac{{g''}}{{2g}} - \frac{{g'^2
}}{{2g^2 }} + \frac{{g'}}{{2rg}} = \frac{g}{{M^4 }}\left( {F_0  -
\Lambda } \right) + \frac{g}{{\phi ^2 }}\frac{{\Lambda _{phys}
}}{{M_P^2 }} ,\end{equation}

\begin{equation}\label{55partA}6\frac{{\phi '^2 }}{{\phi ^2 }} + 2\frac{{g'\phi '}}{{g\phi }} +
4\frac{{\phi '}}{{r\phi }} = \frac{g}{{M^4 }}\left( {F - \Lambda }
\right) + 2\frac{g}{{\phi ^2 }}\frac{{\Lambda _{phys} }}{{M_P^2 }}
,\end{equation}

\begin{equation}\label{66partA}4\frac{{\phi ''}}{\phi } + 6\frac{{\phi '^2 }}{{\phi ^2 }} -
2\frac{{g'\phi '}}{{g\phi }} = \frac{g}{{M^4 }}\left( {F - \Lambda
} \right) + 2\frac{g}{{\phi ^2 }}\frac{{\Lambda _{phys} }}{{M_P^2
}} ,\end{equation} where the prime denotes differentiation $d/dr$.
The constant $\Lambda _{phys}$ represents the physical
four-dimensional cosmological constant, where
\begin{equation}\label{4deinstequationsA}R_{\alpha \beta }^{(4)}  -
\frac{1}{2}g_{\alpha \beta } R^{(4)}  = \frac{{\Lambda _{phys}
}}{{M_P^2 }}g_{\alpha \beta }.\end{equation} In this equation
$R_{\alpha \beta }^{(4)}$, $R^{(4)}$ and $M_P$ are
four-dimensional physical quantities: Ricci tensor, scalar
curvature and Planck scale.

In the case $\Lambda _{phys}=0$  from the equations
(\ref{4dpartA}), (\ref{55partA}) and (\ref{66partA}) we can find
\begin{equation}\label{sfconA} F' + 4\frac{{\phi '}}{\phi }\left( {F - F_0 } \right) = 0,
\end{equation}
\begin{equation}\label{cfconA}g = \frac{{\delta \phi '}}{r} , \    \ \delta=const,
\end{equation}\begin{equation} \label{equationA}r\frac{{\phi ''}}{\phi } + 3r\frac{{\phi '^2 }}{{\phi ^2 }} +
\frac{{\phi '}}{\phi } = \frac{{rg}}{{2M^4 }}\left( {F - \Lambda }
\right) ,\end{equation} where $\delta$ denotes the integration
constant. The (\ref{sfconA}) represents the connection between
source functions, it is simply a consequence of the conservation
of the stress-energy tensor and can be also independently derived
directly from $D_A T_B^A = 0$.

Suppose that the "widht" of brane sitting in the origin $r=0$ is
equal to $\varepsilon$. It is worthwhile to mention that the brane
is assumed to have the nonvanishing "brane width" since the "brane
width" \ \ $\varepsilon$\ \ appears in the later arguments of
localization of the bulk fields and plays a role as a
short-distance cutoff  \cite{Oda1}. Outside a core of radius
$\varepsilon$ we take the source functions in the form $F\left( r
\right) = f\phi ^{ - 2}$, where $f$ is some constant. Taking the
first integral of the last equation \cite{mg1}, we get

\begin{equation}\label{equationAA}r\phi ' =  - \frac{{\delta \Lambda }}{{10M^4 }}\left( {\phi ^2  -
\frac{{5f}}{{3\Lambda }}} \right) + \frac{C}{{\phi ^3 }} ,
\end{equation} where  $C$ is the integration constant. Setting $C=0$,  introducing the positive parameters \begin{equation}a=
\frac{{\delta \Lambda }}{{10M^4 }}  > 0,\ \
d^2=\frac{{5f}}{{3\Lambda }}   > 0, \end{equation}  and imposing
boundary condition at the infinity of transverse space
\begin{equation}\left. {\phi \left( r \right)} \right|_{r =  + \infty }  = const >
0 ,
\end{equation}  we can easily find  two solutions of the equation
(\ref{equationAA}) in the following cases:

\ \ \ \ \ \ \ \ \ \ \ \ \ \ \ \ \ \ \ i) $\Lambda  > 0$, $f > 0$,
$\delta
> 0$
\begin{equation}\label{Solution}\phi \left( r \right) = d\frac{{r^b  - c }}{{r^b  + c
}},\ \ \ g \left( r \right) = 2\delta db\frac{{c r^{b - 2}
}}{{\left( {r^b  + c } \right)^2 }};\end{equation}

\ \ \ \ \ \ \ \ \ \ \ \ \ \ \ \ \ \ \ ii) $\Lambda  < 0$, $f < 0$,
$\delta < 0$
\begin{equation}\label{Solution1}\phi  \left( r \right) = d\frac{{r^b  + c }}{{r^b  - c
}},\ \ \ g \left( r \right) = 2\left| \delta \right|db\frac{{c
r^{b - 2} }}{{\left( {r^b  - c } \right)^2 }};\end{equation} where
$c>0$  is positive integration constant and we have introduced the
parameter $b = 2ad$. The first solution (\ref{Solution}) is the
same one found in \cite{GogberSigleton}. This solution exists in
the case of positive bulk cosmological constant $\Lambda>0$. The
scale factor $\phi(r)$ of this solution is increasing function
asymptotically approaching the finite value at the radial
infinity. The second solution (\ref{Solution1}) is the new one. It
exists for the negative cosmological constant $\Lambda<0$, and its
scale factor $\phi(r)$ is decreasing function asymptotically
approaching the finite non-zero value at the radial infinity. In
this article we consider the second solution (the first one has
been already examined in \cite{GogberSigleton}).

To avoid singularities on the brane (sitting at the origin $r=0$)
we take the boundary conditions for the solution in the form
\begin{equation}\label{boundaryconditions1}\left. {\phi  \left( r \right)} \right|_{r = \varepsilon }  =
1,\ \ \ \left. {\phi  \left( r \right)} \right|_{r =  + \infty } =
d.\end{equation} In (\ref{boundaryconditions1})\ \ $\varepsilon$ \
\ denotes the "brane width", which takes a fixed value. This
boundary conditions allow us to express the integration constant
$c$ in terms of the "brane width":
\begin{equation}\label{c2}c  = \frac{{1 - d}}{{1 + d}}\varepsilon ^b
, \ \ \ {\rm{where }}\ \ 0<d < 1.\end{equation}

Now we turn our attention to the problem of the localization of
the bulk fields on the brane in the background geometry
(\ref{Solution1}). Of course, in due analysis, we will neglect the
back-reaction on the geometry induced by the existence of the bulk
fields, and from now on, without loss of generality, we shall take
a flat metric on the brane.

We start with a massless, spin $0$, real scalar coupled to
gravity:
\begin{equation}\label{RealScalar}S_0   =  - \frac{1}{2}\int {d^6 x\sqrt { - {}^6g} g^{AB}
\partial _A \Phi \partial _B \Phi }.\end{equation} The corresponding equation of
motion has the form
\begin{equation}\label{RealScalarEqMotion}\frac{1}{{\sqrt { - {}^6g} }}\partial _M \left( {\sqrt { - {}^6g}
g^{MN} \partial _N \Phi } \right) = 0.\end{equation}  It turns out
that in the background metric (\ref{Solution1}) the zero-mode
solution of (\ref{RealScalarEqMotion}) is $\Phi _0 \left( {x^M }
\right) = \upsilon \left( {x^\mu  } \right)\rho _0$, where $\rho
_0=constant$,  and $\upsilon \left( {x^\mu  } \right)$ satisfies
the Klein-Gordon equation on the brane $\eta ^{\mu \nu }
\partial _\mu  \partial _\nu  \upsilon \left( {x^\mu  } \right) =
0$. Substituting this solution into the starting action
(\ref{RealScalar}), the action can be cast to
\begin{equation}\label{RealScalarZeroModeAction}S_0  =  - \frac{1}{2}\rho _0^2 \int_0^{2\pi } {d\theta }
\int_\varepsilon ^{ + \infty } {dr\phi ^2 gr} \int {d^4 x\sqrt { -
\eta } \eta ^{\mu \nu } \partial _\mu  \upsilon \left( {x^\alpha }
\right)\partial _\nu  \upsilon \left( {x^\alpha  } \right)}  + ...
=\end{equation}
\begin{equation}\label{RealScalarZeroModeAction1}=  - \pi \delta\rho _0^2 \int_{\phi \left( \varepsilon
 \right)}^{\phi \left( { + \infty } \right)} {\phi ^2 d\phi } \int {d^4 x\sqrt { - \eta } \eta ^
 {\mu \nu } \partial _\mu  \upsilon \left( {x^\alpha  }
\right)\partial _\nu  \upsilon \left( {x^\alpha  } \right)}  +
...= \end{equation}
\begin{equation}\label{RealScalarZeroModeAction1Second} =  - \frac{{\pi |\delta|\rho _0^2
}}{3}\left( {1 - d^3 } \right)\int {d^4 x\sqrt { - \eta } \eta
^{\mu \nu } \partial _\mu  \upsilon \left( {x^\alpha  }
\right)\partial _\nu  \upsilon \left( {x^\alpha  } \right)}  +
...\ \ .\end{equation} The integral over $r$ in
(\ref{RealScalarZeroModeAction}) is finite, so the 4-dimensional
scalar field is localized on the brane.

For spin ${\raise0.7ex\hbox{$1$} \!\mathord{\left/
 {\vphantom {1 2}}\right.\kern-\nulldelimiterspace}
\!\lower0.7ex\hbox{$2$}}$ fermion starting action is the Dirac
action given by
\begin{equation}\label{FermionAction}S_{\frac{1}{2}}  = \int {d^6 x\sqrt { - {}^6g} \overline \Psi
i\Gamma ^M D_M \Psi } ,\end{equation} from which the equation of
motion is given by
\begin{equation}\label{FermionEqMotion}\Gamma ^M D_M \Psi  = \left( {\Gamma ^\mu  D_\mu   + \Gamma ^r D_r
+ \Gamma ^\theta  D_\theta  } \right)\Psi  = 0 \ \ .\end{equation}
We introduce the vielbein $h_A^{\widetilde A}$ through the usual
definition $g_{AB}  = h_A^{\widetilde A} h_B^{\widetilde B} \eta
_{\widetilde A\widetilde B}$ where $\widetilde A,\widetilde B,...$
denote the local Lorentz indices. $\Gamma ^A$  in a curved
space-time is related to $\gamma ^A$ by $\Gamma ^A  =
h_{\widetilde A}^A \gamma ^{\widetilde A}$. The spin connection
$\omega _M^{\widetilde M\widetilde N}$ in the covariant derivative
\ \  $D_M \Psi  = \left( {\partial _M  + \frac{1}{4}\omega
_M^{\widetilde M\widetilde N} \gamma _{\widetilde M\widetilde N} }
\right)\Psi$ \ \ is defined as
$$\omega _M^{\widetilde M\widetilde N}  = \frac{1}{2}h^{N\widetilde M}
\left( {\partial _M h_N^{\widetilde N}  - \partial _N
h_M^{\widetilde N} } \right) - \frac{1}{2}h^{N\widetilde N} \left(
{\partial _M h_N^{\widetilde M}  - \partial _N h_M^{\widetilde M}
} \right) - \frac{1}{2}h^{P\widetilde M} h^{Q\widetilde N} \left(
{\partial _P h_{Q\widetilde R}  - \partial _Q h_{P\widetilde R} }
\right)h_M^{\widetilde R}.$$ After these conventions are set we
can decompose the $6$-dimensional spinor into the form $\Psi
\left( {x^M } \right) = \psi \left( {x^\mu  } \right)A\left( r
\right)\sum {e^{il\theta } } $.  We require that the
four-dimensional part satisfies the massless equation of motion
$\gamma ^\mu  \partial _\mu  \psi \left( {x^\beta  } \right) = 0$
and the chiral condition $\gamma ^r \psi \left( {x^\mu  } \right)
= \psi \left( {x^\mu  } \right)$. As a result we obtain the
following equation for the $s$-wave mode
\begin{equation}\label{FermionZeroModeEq}\left[ {\partial _r  + 2\frac{{\phi '}}{\phi } +
\frac{1}{2}\frac{{\partial _r \left( {rg^{\frac{1}{2}} }
\right)}}{{rg^{\frac{1}{2}} }}} \right]A\left( r \right) = 0.
\end{equation} The solution to this equation reads:
\begin{equation}\label{FermionZeroModeEqSolution}A\left( r \right) = A_0 \phi ^{ - 2} g^{ - \frac{1}{4}} r^{ -
\frac{1}{2}},\end{equation} with $A_0$ being an integration
constant. Substituting this solution into the Dirac action
(\ref{FermionAction}) we have
\begin{equation}\label{FermionActionZeroMode}S_{\frac{1}{2}}  = 2\pi A_0^2 \int_\varepsilon ^{ + \infty }
{dr\phi ^{ - 1} g^{\frac{1}{2}} } \int {d^4 x\sqrt { - \eta }
\overline \psi  i\gamma ^\mu  \partial _\mu  \psi }  + ...
=\end{equation}
\begin{equation}\label{FermionActionZeroModeSecond} = 2\pi A_0^2 \sqrt
{\frac{{2\left| \delta  \right|bc }}{d}} \int_\varepsilon ^{ +
\infty } {\frac{{r^{\frac{1}{2}b} dr}}{{r\left( {r^b  + c }
\right)}}} \int {d^4 x\sqrt { - \eta } \overline \psi  i\gamma
^\mu  \partial _\mu  \psi }  + ... \ \ .\end{equation}
 As long as
$b>0$  and the "brane width" $\varepsilon$ is nonvanishing the
integral over $r$ is obviously finite. Indeed, the integrand in
(\ref{FermionActionZeroModeSecond}) scales as $r^{ - \frac{1}{2}b
- 1}$ at the radial infinity and is the smooth functions between
$r=\varepsilon$ and $r =  + \infty$, so this integral over $r$ is
finite even if the analytic expression is not available. So the
spin ${\raise0.7ex\hbox{$1$} \!\mathord{\left/
 {\vphantom {1 2}}\right.\kern-\nulldelimiterspace}
\!\lower0.7ex\hbox{$2$}}$  fermion is localized on the brane only
by the gravitational interaction.

Now let us consider  the action of $U(1)$  vector field:
\begin{equation}\label{VectorFieldAction}S_1  =  - \frac{1}{4}\int {d^6 x\sqrt { - {}^6g} g^{AB} g^{MN}
F_{AM} F_{BN} } ,\end{equation} where $F_{MN}  = \partial _M A_N -
\partial _N A_M$  as usual. From this action the equation of
motion is given by
\begin{equation}\label{VectorFieldEqMotion}\frac{1}{{\sqrt { - {}^6g} }}\partial _M \left( {\sqrt { - {}^6g}
g^{MN} g^{RS} F_{NS} } \right) = 0.\end{equation} By choosing the
gauge condition $A_{\theta}=0$ and decomposing the vector field as
\begin{equation}\label{decomposition1}A_\mu  \left( {x^M } \right) = a_\mu  \left( {x^\mu  }
\right)\sum\limits_{l,m} {\sigma _m \left( r \right)e^{il\theta }
},\end{equation}
 \begin{equation}\label{decomposition2}A_r \left( {x^M } \right) = a_r \left( {x^\mu  }
\right)\sum\limits_{l,m} {\sigma _m \left( r \right)e^{il\theta }
},\end{equation} it is straightforward to see that there is the
$s$-wave $(l=0)$ constant  solution $\sigma _m \left( r
\right)=\sigma _0=const$ and  $a_r=const$.  In deriving this
solution we have used $\partial _\mu  a^\mu   = \partial ^\mu
f_{\mu \nu }  = 0$ with the definition of $f_{\mu \nu }  =
\partial _\mu  a_\nu   - \partial _\nu  a_\mu$. As in the previous
cases, let us substitute this constant solution into the action
(\ref{VectorFieldAction}). It turns out that the action is reduced
to
\begin{equation}\label{VectorFieldActionZeroMode}S_1  =  - \frac{\pi }{2}\sigma _0^2 \int_\varepsilon ^{ + \infty }
{drgr} \int {d^4 x\sqrt { - \eta } \eta ^{\alpha \beta } \eta
^{\mu \nu } f_{\alpha \mu } f_{\beta \nu } }+... =\end{equation}
\begin{equation}\label{VectorFieldActionZeroModeSecond}= \frac{{\pi \left| \delta
\right|}}{2}\sigma _0^2 \int_{\phi  \left( \varepsilon
\right)}^{\phi  \left( { + \infty } \right)} {d\phi  } \int {d^4
x\sqrt { - \eta } \eta ^{\alpha \beta } \eta ^{\mu \nu } f_{\alpha
\mu } f_{\beta \nu } }  + ... =  - \frac{{\pi \left| \delta
\right|}}{2}\sigma _0^2 \left( {1 - d} \right)\int {d^4 x\sqrt { -
\eta } \eta ^{\alpha \beta } \eta ^{\mu \nu } f_{\alpha \mu }
f_{\beta \nu } }  + ... \ \ .\end{equation}

 As we can see
from (\ref{VectorFieldActionZeroModeSecond}) the integral over $r$
in (\ref{VectorFieldActionZeroMode}) is finite. Thus, the vector
field is localized on the brane.

Next we consider spin ${\raise0.7ex\hbox{$3$} \!\mathord{\left/
 {\vphantom {3 2}}\right.\kern-\nulldelimiterspace}
\!\lower0.7ex\hbox{$2$}}$ field (the gravitino). We begin with the
action of the Rarita-Schwinger gravitino
\begin{equation}\label{GravitinoAction}S_{\frac{3}{2}}  = \int {d^6 x\sqrt { - {}^6g} \overline \Psi  _A
i\Gamma ^{\left[ A \right.} \Gamma ^B \Gamma ^{\left. C \right]}
D_B \Psi _C },\end{equation}from which the equation of motion is
given by
\begin{equation}\label{GravitinoEqMotion}\Gamma ^{\left[ A \right.} \Gamma ^B \Gamma ^{\left. C \right]}
D_B \Psi _C  = 0.\end{equation} Here the square bracket denotes
the anti-symmetrization and the covariant derivative is defined
with the affine connection $\Gamma _{BC}^A  = h_{\widetilde B}^A
\left( {\partial _B h_C^{\widetilde B}  + \omega _B^{\widetilde
B\widetilde C} h_{C\widetilde C} } \right)$  by
\begin{equation}\label{Derivative}D_A \Psi _B  = \partial _A \Psi _B  - \Gamma _{AB}^C \Psi _C  +
\frac{1}{4}\omega _A^{\widetilde A\widetilde B} \gamma
_{\widetilde A\widetilde B} \Psi _B\end{equation} After taking the
gauge condition $\Psi _\theta   = 0$ we look for the solutions of
the form  $\Psi _\mu  \left( {x^A } \right) = \psi _\mu  \left(
{x^\nu  } \right)u\left( r \right)\sum {e^{il\theta } }$, $\Psi _r
\left( {x^A } \right) = \psi _r \left( {x^\nu  } \right)u\left( r
\right)\sum {e^{il\theta } }$ where  $\psi _\mu  \left( {x^\nu  }
\right)$  satisfies the following equations $\gamma ^\nu  \psi
_\nu   =
\partial ^\mu  \psi _\mu   = \gamma ^{\left[ \nu  \right.} \gamma
^\rho  \gamma ^{\left. \tau \right]}
\partial _\rho  \psi _\tau   = 0$, $\gamma ^r \psi _\nu   = \psi
_\nu$. For the $s$-wave solution and $\psi _r \left( {x^\nu  }
\right)=0$ we have for the equation of motion
(\ref{GravitinoEqMotion}) the following form
\begin{equation}\label{GravitinoEqMotionZeroMode}\left[ {\partial _r  + \frac{3}{2}\frac{{\phi '}}{\phi } +
\frac{1}{2}\frac{{\partial _r \left( {rg^{\frac{1}{2}} }
\right)}}{{rg^{\frac{1}{2}} }}} \right]u\left( r \right) =
0.\end{equation} The solution to this equation is
\begin{equation}\label{GravitinoZeroMode}u\left( r \right) = u_0 \phi ^{ - \frac{3}{2}} g^{ - \frac{1}{4}}
r^{ - \frac{1}{2}},\end{equation} with $u_0$  being an integration
constant. Substituting this solution into the action
(\ref{GravitinoAction}) we get
\begin{equation}\label{GravitinoActionZeroMode}S_{\frac{3}{2}}  = 2\pi u_0^2 \int_\varepsilon ^{ + \infty }
{dr\phi ^{ - 2} g^{\frac{1}{2}} } \int {d^4 x\sqrt { - \eta }
\overline \psi  _\mu  i\gamma ^{\left[ \mu  \right.} \gamma ^\nu
\gamma ^{\left. \rho  \right]} \partial _\nu  \psi _\rho  }  +
...=\end{equation}
\begin{equation}\label{GravitinoActionZeroModeSecond}= 2\pi u_0^2 \sqrt
{\frac{{2\left| \delta  \right|bc_2 }}{d}} \int_\varepsilon ^{ +
\infty } {\frac{{r^{\frac{1}{2}b} \left( {r^b  - c_2 }
\right)}}{{r\left( {r^b  + c_2 } \right)^2 }}dr} \int {d^4 x\sqrt
{ - \eta } \overline \psi  _\mu  i\gamma ^{\left[ \mu  \right.}
\gamma ^\nu  \gamma ^{\left. \rho  \right]} \partial _\nu  \psi
_\rho  }  + ... \ \ .\end{equation}
 As in the case of spin
${\raise0.7ex\hbox{$1$} \!\mathord{\left/
 {\vphantom {1 2}}\right.\kern-\nulldelimiterspace}
\!\lower0.7ex\hbox{$2$}}$ fermion
(\ref{FermionActionZeroModeSecond}) the integrand in
(\ref{GravitinoActionZeroModeSecond}) scales as $ r^{ -
\frac{1}{2}b - 1} $ at the radial infinity and is smooth function
between $r=\varepsilon$ and $ r = + \infty$, so the integral over
$r$ is finite as long as the "brane width" $\varepsilon$ is
non-zero. This means that spin ${\raise0.7ex\hbox{$3$}
\!\mathord{\left/
 {\vphantom {3 2}}\right.\kern-\nulldelimiterspace}
\!\lower0.7ex\hbox{$2$}}$ field is localized on the brane.

Now let us consider spin $2$ gravitational field. In this case we
consider the spin-$2$ metric fluctuations $H_{\mu \nu }$:
\begin{equation}\label{MetricFluctuation}ds^2  = \left\{ {\phi ^2 \left( r \right)\eta _{\alpha \beta }
\left( {x^\nu  } \right) + H_{\alpha \beta } } \right\}dx^\alpha
dx^\beta   - g\left( r \right)\left( {dr^2  + r^2 d\theta ^2 }
\right).\end{equation} The corresponding equation of motion for
the fluctuations has the following form:
\begin{equation}\label{MetricFluctuationEqMotion}\frac{1}{{\sqrt { - {}^6g} }}\partial _A \left( {\sqrt { - {}^6g}
g^{AB} \partial _B H_{\mu \nu } } \right) = 0.\end{equation} From
the 4-dimensional point of view fluctuations are described by a
tensor field $H_{\mu \nu }$ which is transverse and traceless:
\begin{equation}\partial _\mu  H_\nu ^\mu   = 0,\ \ H_\mu ^\mu   =
0.
\end{equation}
This tensor is invariant under 4-dimensional general coordinate
transformations. We look for solutions of the form
\begin{equation} \label{MetricFluctuation1}H_{\alpha \beta } =
h_{\alpha \beta } (x^\nu )\sum\limits_{ml} {\tau _m (r)\exp
(i\theta l)},
\end{equation} where $\partial ^2 h_{\mu \nu } (x^\alpha  ) = m_0^2 h_{\mu \nu } (x^\alpha
)$. It is easy to show that the equation of motion
(\ref{MetricFluctuationEqMotion}) has the zero-mass ($m_0=0$) and
$s$-wave ($l=0$) constant solution $\tau_0=const$. Substitution of
this zero mode into the Einstein-Hilbert action leads to
\begin{equation}\label{EinsteinHilbertAction}S_2  \sim 2\pi \tau _0^2 \int_\varepsilon ^{ + \infty } {dr\phi ^2
gr} \int {d^4 x\left[ {\partial ^\rho  h^{\alpha \beta } \partial
_\rho  h_{\alpha \beta }  + ...} \right]}=\end{equation}
\begin{equation}\label{EinsteinHilbertAction1}
 = 2\pi \delta \tau _0^2 \int_{\phi \left( \varepsilon  \right)}^{\phi \left( {
 + \infty } \right)} {\phi ^2 d\phi } \int {d^4 x\left[ {\partial ^\rho  h^{\alpha
 \beta } \partial _\rho  h_{\alpha \beta }  + ...} \right]}  = \frac{2}{3}\pi \left|
  \delta  \right|\tau _0^2 \left( {1 - d^3 } \right)\int {d^4 x\left[ {\partial ^\rho
   h^{\alpha \beta } \partial _\rho  h_{\alpha \beta }  + ...} \right]}
.\end{equation} The integral over $r$ is finite, so the bulk
graviton is trapped on the brane. With our background metric the
general expression for the four-dimensional Planck scale $M_P$,
expressed in terms of $M$, is
\begin{equation}\label{PlanckScale}M_P^2  = 2\pi M^4 \int_\varepsilon ^{ + \infty } {dr\phi ^2 gr}  =
2\pi \delta M^4 \int_\varepsilon ^{ + \infty } {dr\phi ^2 \phi '}
=  \frac{2}{3}\pi |\delta| M^4 (1-d^3).\end{equation}  The
inequality $ M \ll M_P $ is possible by adjusting $ M^4 $ and the
product $ \left| \delta  \right|(1 - d^3 ) $, and thus could lead
to a solution of the gauge hierarchy problem.

Finally, let us consider the totally antisymmetric tensor fields.
The action of $k$-rank totally antisymmetric tensor field $A_k$ is
of the form
\begin{equation}\label{TotallyAntisymmetricFieldAction}S_k  =  -
\frac{1}{2}\int {F_{k + 1}  \wedge  * F_{k + 1} ,}
\end{equation} where $ F_{k + 1}  = dA_k $.
The corresponding equation of motion is given by
\begin{equation}\label{TotallyAntisymmetricFielEqMotion} d \wedge  * F_{k + 1}
 = 0 . \end{equation} It is easy to show that  $A_{\mu _1 \mu _2 ...\mu _k }  = a_{\mu _1 \mu _2 ...\mu _k }
\left( {x^\nu  } \right)u_0$  with $u_0=const$ is a solution to
the equation of motion (\ref{TotallyAntisymmetricFielEqMotion}) if
$ d \wedge * f = 0$ where $ f = da$. Substituting this solution in
the action (\ref{TotallyAntisymmetricFieldAction}) leads to the
expression
\begin{equation}\label{TotallyAntisymmetricFieldActionZeroMode1}
S_k  \sim \int_\varepsilon ^{ + \infty } {dr\phi ^{2 - 2k} gr}
\int {f_{k + 1}  \wedge  * f_{k + 1}  + ...}  =
\end{equation}
\begin{equation}\label{TotallyAntisymmetricFieldActionZeroMode2}
 = \delta \int_{\phi \left( \varepsilon  \right)}^{\phi \left( { +
 \infty } \right)} {\phi ^{2 - 2k} d\phi } \int {f_{k + 1}  \wedge  * f_{k + 1}  + ...}
 = \frac{{\left| \delta  \right|}}{{\left( {3 - 2k} \right)}}\left( {1 - d^{3 - 2k} }
 \right)\int {f_{k + 1}  \wedge  * f_{k + 1}  + ...}
\end{equation} As we can see from
(\ref{TotallyAntisymmetricFieldActionZeroMode2}) the integral over
$r$ in (\ref{TotallyAntisymmetricFieldActionZeroMode1}) is finite,
so the totally antisymmetric tensor fields are also localized on
the brane by the gravitational interaction.

In conclusion, in this article we have presented a new $3$-brane
solution with a decreasing scale factor $\phi \left( r \right)$.
This solution is found for the negative bulk cosmological constant
$\Lambda  < 0$.  In addition we have presented a complete analysis
of localization of a bulk fields on a brane via the gravitational
interaction without including an additional interection. The
technical reason of the localization of all bulk fields lies not
only in the fact that the scale factor $\phi \left( r \right)$  is
a smooth function without singularities from the edge of the brane
to the radial infinity and approaches a definite non-zero value at
the infinity. The main reason is connected with the properties of
the function $g(r)$.  On the edge of the brane ($r=\varepsilon$)
it has a value $ g\left( \varepsilon \right) = a\delta
(1-d^2)\varepsilon^{-2}$ and fast enough tends to zero as one
moves of the brane.  In the case $\varepsilon \ll 1$, $\delta  \gg
1$ and $b=2ad=\frac{{\delta \Lambda }}{{5M^4 }}d \gg 1 $ this
function has a $\delta$-like behavior, and in contrast to the
model considered in \cite{Oda1} in our model the wave functions of
the zero-mode solutions of the bulk fields along the extra
dimensions are peaked at the location of the brane.

{\bf Acknowledgements:} P.M. would like to acknowledge the
hospitality extended during his visits at the Abdus Salam
International Centre for Theoretical Physics where the main part
of this work was done.

\end{document}